\documentstyle[11pt]{article}

\def\sect{\section}
\def\EQ{\begin{equation}}
\def\EN{\end{equation}}
\def\bea{\begin{eqnarray}}
\def\ena{\end{eqnarray}}
\newcommand{\vs}[1]{\vspace{#1 mm}}

\begin{document}
\begin{titlepage}
\begin{center}

\hfill  {\sc Pre-Print\ HUTP-98/A034} \\
\hfill  {\sc Pre-Print\ NBI-HE-98-09} \\
\hfill  {\sc Pre-Print\ MIT-CTP$\sharp$2729} \\

\hfill  {\sc April 1998} \\
\hfill  {\tt hep-th/9805074} \\
        [.2in]

{\large\bf
HETEROTIC T-DUALITY AND THE RENORMALIZATION GROUP
}\\[.3in]

{\bf Kasper Olsen}\footnote{E-mail: {\tt kolsen@lyman.harvard.edu}} \\
{\it Lyman Laboratory of Physics, Harvard University}\\
{\it Cambridge, MA 02138, U.S.A.}\\
and\\
{\it The Niels Bohr Institute, Blegdamsvej 17}\\
{\it DK-2100 Copenhagen, Denmark}\\

\bigskip

{\bf Ricardo Schiappa}\footnote{E-mail: {\tt ricardos@ctp.mit.edu}} \\
{\it Center for Theoretical Physics and Department of Physics}\\
{\it Massachusetts Institute of Technology, 77 Massachusetts Ave.}\\
{\it Cambridge, MA 02139, U.S.A.}\\

\end{center}

\vs{5}
\centerline{{\bf{Abstract}}}
\vs{5}

We consider target space duality transformations for heterotic sigma models 
and strings away from renormalization group fixed points. By imposing 
certain consistency requirements between the $T$-duality symmetry and 
renormalization group flows, the one loop gauge beta function is uniquely 
determined, without any diagram calculations. Classical $T$-duality symmetry 
is a valid quantum symmetry of the heterotic sigma model, severely 
constraining its renormalization flows at this one loop order. The issue of 
heterotic anomalies and their cancelation is addressed from this duality 
constraining viewpoint.

\vfill



\end{titlepage}

\newpage
\renewcommand{\thefootnote}{\arabic{footnote}}
\setcounter{footnote}{0}

\sect{Introduction}

Symmetry is a central concept in quantum field theory. Usually, one 
thinks of symmetries as transformations acting on the fields of a 
theory, leaving its partition function invariant. More fashionable 
these days is a different concept. This is the idea of duality symmetries, 
transformations on the parameter space of a theory which leave its 
partition function invariant. One such example is the well known 
target space duality ($T$-duality henceforth), see \cite{porra} for 
a complete set of references. Another important action on the parameter 
space of a quantum field theory is that of the renormalization group 
(RG henceforth), as RG transformations also leave the partition 
function invariant. The study of the interplay between the RG and 
duality symmetries then seems quite natural \cite{lut}.

The idea of $T$-duality symmetry first came about in the context of string 
theory \cite{porra}, but it was soon realized that a proof of its 
existence could be given directly from sigma model path integral 
considerations \cite{bus1,bus2}. On the other hand, sigma models are 
well defined two dimensional quantum field theories away from the 
conformal backgrounds that are of interest for string theory 
\cite{fried,freed}. A study was then initiated concerning the possibility 
of having $T$-duality as a symmetry of the quantum sigma model away from 
the (conformal) RG fixed points, when the target manifold admits an 
Abelian isometry. Central to this study was the aforementioned interplay 
between the duality symmetry and the RG . It was observed that this 
interplay translates to consistency conditions to be verified by the RG 
flows of the model; and that indeed they were verified by, and only by, 
the correct RG flows of the bosonic sigma model. Such a study was carried 
out up to two loops, order ${\cal O}({\alpha'}^2)$, in references 
\cite{haag,HO,HOS,haag2}.

Such symmetry being verified in the bosonic sigma model -- where the 
target fields are a metric, an antisymmetric field and a dilaton -- one 
then wonders what happens for the supersymmetric extensions of such 
models. With relation to the ${\cal N}=2$ supersymmetric sigma model
\cite{freed}, where the target fields are similar, the bosonic results 
do have something to say. This is due to the fact that this model asks for 
target K\"ahler geometry \cite{susy1,susy2}, if it is to be supersymmetric. 
Including this extra constraint in the analysis of \cite{haag,HO,HOS} one 
then sees that when restricted to background K\"ahler tensor structures, 
the results obtained in there translate to the well known results for this 
supersymmetric sigma model. Corresponding results for the ${\cal N}=1$ 
supersymmetric sigma model can also be obtained.

Another interesting supersymmetric extension of the bosonic sigma 
model is the heterotic sigma model \cite{hullwitt}. One extra feature 
is that one now has a target gauge field. It is this new 
coupling that we shall study in here, following the point of view 
in \cite{haag,HO,HOS,haag2}. We shall work to one loop, order 
${\cal O}(\alpha')$, and we will see that $T$-duality is again a good 
quantum symmetry of this sigma model. This shall be done by deriving 
consistency conditions for the RG flows of the model under $T$-duality 
and observing that they are satisfied by, and only by, the correct 
RG flows of the heterotic sigma model. However, yet another extra 
feature arises. In these models the measure of integration 
over the quantum fields involves chiral fermions. Such fermions 
produce potential anomalies, and we therefore have a first example 
where we can analyze the interplay of $T$-duality and the RG flow 
in the presence of anomalies. It is then reasonable to expect that 
the consistency conditions may have something to say about these 
anomalies, as they need to cancel in order to define an RG flow.

One should finally remark that it is indeed interesting that 
duality, a symmetry which is apparently entirely unaware of the 
renormalization structure of the model, should yield such strong 
constraints as to uniquely determine the sigma model beta functions. 
Work similar in spirit to the one we perform here has also been 
carried out in condensed matter systems \cite{cond1,cond2}, and more 
recently in systems that have a strong-weak coupling duality \cite{sw1,sw2}.

Following \cite{haag,HO,HOS,haag2}, let us begin with a theory with an 
arbitrary number of couplings, $g^i$, $i=1,\dots,n$, and consider 
a duality symmetry, $T$, acting as a map between equivalent points 
in the parameter space, such that,
$$
Tg^i \equiv \tilde{g}^i = \tilde{g}^i(g). 	\eqno(1.1)
$$
Let us also assume that our system has a renormalization group flow, $R$, 
encoded by a set of beta functions, and acting on the parameter space by,
$$
Rg^i \equiv \beta^i(g) = \mu{d g^i \over d\mu},		\eqno(1.2)
$$
where $\mu$ is some appropriate subtraction scale. Given any function 
on the parameter space of the theory, $F(g)$, the previous operations 
act as follows:
$$
TF(g)=F(\tilde{g}(g)) \qquad , \qquad
RF(g)={\delta F \over \delta g^i}(g)  \cdot  \beta^i(g).	\eqno(1.3)
$$
For a finite number of couplings the derivatives above should be understood 
as ordinary derivatives, whereas in the case of the sigma model these will 
be functional derivatives, and the dot will imply an integration over the 
target manifold.

The consistency requirements governing the interplay of the duality symmetry 
and the RG can now be stated simply as,
$$
[T,R]=0,	\eqno(1.4)
$$
or in words: duality transformations and RG flows commute as motions in the 
parameter space of the theory. This amounts to a set of consistency 
conditions on the beta functions of our system:
$$
\beta^i(\tilde{g})={\delta \tilde{g}^i \over \delta g^j} \cdot 
\beta^j(g).	\eqno(1.5)
$$
As we shall see, this is a very strict set of requirements in our model.

The organization of this paper is as follows. In section 2 we will give 
a brief review of the heterotic sigma model, and on how to construct the 
$T$-duality transformation acting on the target space. Then, in section 3, 
we shall see how these transformations translate to a set of consistency 
conditions to be satisfied by the beta functions of the model. In section 4, 
we study such conditions in the heterotic sigma model; for the simpler case 
of torsionless backgrounds and paying special attention to the cancelation 
of anomalies. The results obtained in this section are then extended to 
torsionfull backgrounds in section 5, where the calculations are more 
involved. Finally, in section 6, we present a concluding outline.

\sect{Duality in the Heterotic Sigma Model}

We shall start by reviewing the construction of the heterotic sigma 
model in $(1,0)$ superspace, and the standard procedure of dualizing such 
model. We will closely follow the main references on the subject, 
\cite{hullwitt,hull,alv1,alv2}, and refer to them for further details.

Superspace will have two bosonic coordinates, $z_0$ and $z_1$, and 
a single fermionic coordinate of positive chirality, $\theta$. The 
supersymmetry is ${\cal N}={1 \over 2}$ Majorana-Weyl, as only the 
left moving bosons have fermionic partners. We will 
consider two types of superfields, one scalar coordinate superfield 
and one spinor gauge superfield,
$$
\Phi^{\mu}(z,\theta) = X^{\mu}(z) + \theta\lambda^{\mu}(z) 
\qquad , \qquad 
\Psi^{I}(z,\theta) = \psi^{I}(z) + \theta F^{I}(z).	\eqno(2.1)
$$
In here the $\Phi^\mu$ are coordinates in a $(d+1)$-dimensional 
target manifold ${\cal M}$, so that $\mu=0,1,...,d$, while the 
$\Psi^I$ are sections of a $G$-bundle over ${\cal M}$ with $n$-dimensional 
fibers, so that $I=1,...,n$. 
These spinor superfields transform under a representation 
$R$ of the gauge group $G$, with $n={\rm dim}\,R$. We will consider 
arbitrary $n$, $d$, even though for the heterotic 
superstring $d+1=10$, $n=32$ and $G={\rm Spin}(32)
/{\bf Z}_2$ or $G=E_8\otimes E_8$ \cite{gross}.
Using light-cone coordinates, $z^{\pm}={1 \over \sqrt 2}(z^0\pm z^1)$ 
and $\partial_\pm={1 \over \sqrt 2}(\partial_0\pm\partial_1)$, the 
superspace $(1,0)$ covariant derivative is written as:
$$
D = \partial_\theta + i\theta\partial_+\qquad ,\qquad D^2=i\partial_+. 
\eqno(2.2)
$$

We consider the target manifold endowed with a metric $g_{\mu\nu}$, 
antisymmetric tensor field $b_{\mu\nu}$ and a gauge connection 
${A_\mu}_{IJ}$ associated to the gauge group $G$. The Lagrangian density 
of such model is given by \cite{hullwitt,alv1}:
$$
{\cal L} = -i \int d\theta \, \{ \,
\bigl( g_{\mu\nu}(\Phi) + b_{\mu\nu}(\Phi) \bigr)
D\Phi^{\mu}\partial_-\Phi^{\nu}
-i \delta_{IJ} \Psi^I 
\bigl( D\Psi^J + {{A_\mu}^J}_K(\Phi)D\Phi^{\mu}\Psi^K  \bigr)
\, \}.	\eqno(2.3)
$$
One should keep in mind that the action has an overall coefficient of 
${1 \over 4\pi\alpha'}$, as usual. A good exercise is to do the $\theta$ 
integration and eliminate the auxiliary fields. One should find:
$$
{\cal L} = 
\bigl( g_{\mu\nu} + b_{\mu\nu} \bigr)
\partial_+X^{\mu}\partial_-X^{\nu} + 
ig_{\mu\nu} \lambda^{\mu} 
\bigl( \partial_-\lambda^{\nu} + 
\bigl( \Gamma^{\nu}_{\rho\sigma}+{1 \over 2}{H^{\nu}}_{\rho\sigma} 
\bigr)\partial_-X^{\rho}\lambda^{\sigma} \bigr) + 
$$
$$
+ i\psi^I \bigl(
\partial_+\psi^I + {{A_{\mu}}^I}_J\partial_+X^{\mu}\psi^J \bigr) + 
{1 \over 2} {F_{\mu\nu}}_{IJ}\lambda^{\mu}\lambda^{\nu}\psi^I\psi^J, 
\eqno(2.4)
$$
where,
$$
H_{\mu\nu\rho}=\partial_{\mu}b_{\nu\rho}+\partial_{\nu}b_{\rho\mu}+
\partial_{\rho}b_{\mu\nu} \qquad {\rm and} \qquad 
F_{\mu\nu}=\partial_\mu A_\nu - \partial_\nu A_\mu +[A_\mu,A_\nu]. 
\eqno(2.5)
$$

We need to assume that the sigma model has an Abelian isometry in the 
target manifold, which will enable duality transformations 
\cite{hull,alv1,alv2}. Let $\xi$ be the Killing vector that generates 
the Abelian isometry. The diffeomorphism generated by $\xi$ transforms 
the scalar superfields, and the total action is invariant under the 
isometry only if we can compensate this transformation in the scalar 
superfields by a gauge transformation in the spinor superfields 
\cite{hull,alv1}. This introduces a target gauge transformation parameter 
$\kappa$, such that 
$\delta_\xi A_\mu\equiv{\pounds}_\xi A_\mu={\cal D}_\mu\kappa$.

Choose adapted coordinates to the Killing vector, 
$\xi^\mu\partial_\mu\equiv\partial_0$, and split the coordinates as
$\mu,\nu=0,1,...,d=0,i$, so that the $\mu=0$ component is singled out. In 
these adapted coordinates the isometry is manifest through independence of 
the background fields on the coordinate $X^0$. Moreover, in these coordinates 
the target gauge transformation parameter will satisfy \cite{hull,alv1},
$$
{\cal D}_{\mu}\kappa \equiv \partial_\mu\kappa+[A_\mu,\kappa]=0. 
\eqno(2.6)
$$

The duality transformations are then \cite{alv1,alv2}:
$$
\tilde{g}_{00} = {1 \over g_{00}} \quad , \quad 
\tilde{g}_{0i} = {b_{0i} \over g_{00}} \quad , \quad 
\tilde{b}_{0i} = {g_{0i} \over g_{00}},
$$
$$
\tilde{g}_{ij} = g_{ij} - {{g_{0i}g_{0j}-b_{0i}b_{0j}}\over{g_{00}}} 
\quad , \quad
\tilde{b}_{ij} = b_{ij} - {{g_{0i}b_{0j}-g_{0j}b_{0i}}\over{g_{00}}}, 
\eqno(2.7)
$$
$$
\tilde{A_0}_{IJ} = -{1 \over g_{00}}\mu_{IJ}, \eqno(2.8)
$$
$$
\tilde{A_i}_{IJ} = {A_i}_{IJ}-{g_{0i}+b_{0i} \over g_{00}}\mu_{IJ}. 
\eqno(2.9)
$$
where we have used $\mu_{IJ}\equiv(\kappa-\xi^\alpha A_\alpha)_{IJ}$ 
following \cite{hull,alv1}, and which in adapted coordinates becomes 
$\mu_{IJ}\equiv(\kappa-A_0)_{IJ}$. Observe that the gauge transformation 
properties of $\kappa$ are such that $\mu_{IJ}$ will transform covariantly 
under gauge transformations \cite{hull,alv1}. Equations (2.7) are 
well known since \cite{bus1,bus2}, and their interplay with the RG has 
been studied in \cite{haag,HO,HOS,haag2}. They shall not be dealt with 
in here, as to our one loop order there is nothing new to be found relative 
to the work in \cite{haag}. We shall rather concentrate on the new additions 
(2.8) and (2.9) yielding the duality transformations for the gauge connection.

There is one more duality transformation one needs to pay attention to, the 
one for the dilaton field. As is well known, in a curved world-sheet 
we have to include one further coupling in our action,
$$
{1 \over 4\pi}\int d^2z \, {\sqrt{h}} R^{(2)}\phi(X),	\eqno(2.10)
$$
where $h=\det h_{ab}$, $h_{ab}$ being the two dimensional world-sheet 
metric, and $R^{(2)}$ its scalar curvature. $\phi(X)$ is the background 
dilaton field in ${\cal M}$. This term is required in order to construct the 
Weyl anomaly coefficients (see section 3). We should however point out 
that the addition of such coupling to the heterotic string is not entirely 
trivial as it is not invariant under the so-called kappa-symmetry 
\cite{bell,bell2}. Taking into account the one loop Jacobian from 
integrating out auxiliary fields in the dualization procedure, one finds 
as usual the dilaton shift \cite{bus2,berg1}:
$$
\tilde{\phi} = \phi - {1 \over 2} \ln g_{00}.		\eqno(2.11)
$$

Formulas (2.7-9) were obtained using classical manipulations alone. Only 
(2.11) involves quantum considerations. So, for this heterotic sigma model, 
we need to be careful in the following as there will be anomalies generated 
by the chiral fermion rotations in the quantum measure, and if so the 
original and dual action will not be equivalent. If we want these two 
theories to be equivalent one must find certain conditions on the background 
fields in order to cancel the anomalies. We shall see in the following that 
the consistency conditions (1.5) do have something to say on this matter.

\sect{Renormalization and Consistency Conditions}

The renormalization of the heterotic sigma model has been studied in many 
references. Of particular interest to our investigations are the one loop 
beta functions \cite{zano,sen,grig,calla}. However, there are some 
subtleties we should point out before proceeding, as the one loop 
effective action is not gauge or Lorentz invariant. It happens that 
this non-invariance is of a very special kind, organizing itself into 
the well known gauge and Lorentz Chern-Simons (order ${\cal O}(\alpha')$) 
completion of the torsion \cite{hullwitt}. Then, starting at two loops, 
there are non-trivial anomalous contributions to the primitive divergences 
of the theory, and things get more complicated \cite{grig}. None of these 
problems will be of concern to us to the order ${\cal O}(\alpha')$ we shall 
be working to, appearing only at order ${\cal O}(\alpha'^2)$. The one loop, 
order ${\cal O}(\alpha')$, beta functions can be computed to be 
\cite{grig,calla}:
$$
\beta^{g}_{\mu\nu} = R_{\mu\nu}-
{1\over4}{H_{\mu}}^{\lambda\rho}H_{\lambda\rho\nu}+
{\cal O}(\alpha'),	\eqno(3.1)
$$
$$
\beta^{b}_{\mu\nu} = -{1\over2}\nabla^{\lambda}H_{\lambda\mu\nu}+
{\cal O}(\alpha'),	\eqno(3.2)
$$
$$
\beta^{A}_{\mu} = {1\over2}({\bf D}^{\lambda}F_{\lambda\mu}+
{1\over2}{H_{\mu}}^{\lambda\rho}F_{\lambda\rho})+
{\cal O}(\alpha'),	\eqno(3.3)
$$
where $R_{\mu\nu}$ is the Ricci tensor of the target manifold, 
$\nabla_\mu$ is the metric covariant derivative, and ${\bf D}_\mu$ 
is the covariant derivative involving both the gauge and the 
metric connections.

Of special interest to us are the Weyl anomaly coefficients 
\cite{tsey1,tsey2,tsey3,bell}, which are in general different from the 
RG beta functions. Their importance comes from the fact that while the 
definition of the sigma model beta functions ($\beta$) is ambiguous due 
to the freedom of target reparameterization, there is no such ambiguity 
for the Weyl anomaly coefficients ($\bar{\beta}$) which are invariant 
under such transformations. This, of course, is related to the fact that 
the $\bar{\beta}$-functions are used to compute the Weyl anomaly, while 
the $\beta$-functions are used to compute the scale anomaly \cite{tsey1,tsey2}.

The advantage of using Weyl anomaly coefficients in our studies is then 
due to the fact that while both $\bar{\beta}$ and $\beta$ satisfy the 
consistency conditions (1.5), the $\bar{\beta}$-functions satisfy them 
exactly, while the $\beta$-functions satisfy them up to a target 
reparameterization \cite{haag,HO}. Since both encode essentially the same 
RG information, in the following we shall simply consider RG motions as 
generated by the $\bar{\beta}$-functions. For the heterotic sigma model 
\cite{tsey3,bell}, and for the loop orders considered in this work:
$$
\bar{\beta}^{g}_{\mu\nu}=\beta^g_{\mu\nu}+2\nabla_\mu\partial_\nu\phi
+{\cal O}(\alpha'),	\eqno(3.4) 
$$
$$
\bar{\beta}^{b}_{\mu\nu}=\beta^b_{\mu\nu}+{H_{\mu\nu}}^\lambda
\partial_\lambda\phi+{\cal O}(\alpha'),		\eqno(3.5)
$$
$$
\bar{\beta}^{A}_\mu=\beta^A_\mu+{F_\mu}^\lambda\partial_\lambda\phi
+{\cal O}(\alpha').	\eqno(3.6)
$$
The consistency conditions (1.5) can now be derived. The couplings are 
$g^i \equiv \{g_{\mu\nu},b_{\mu\nu},A_\mu,\phi \}$, and the duality 
operation (1.1) is defined through (2.7-9) and (2.11). The RG flow 
operation is defined in (1.2), for our couplings, with the only difference 
that we shall consider $\bar{\beta}$-generated RG motions as previously 
explained. It is then a straightforward exercise to write down the 
consistency conditions (1.5) for the heterotic sigma model. The consistency 
conditions associated to (2.7) and (2.11) have in fact been studied 
before \cite{haag,HO,HOS} and are known to be satisfied by, and only by, 
(3.1-2) or (3.4-5). So, we shall not deal with them in here. The consistency 
conditions associated to the gauge field coupling are:
$$
{\bar{\beta}}^{\tilde{A}}_{0} = {1 \over g_{00}}{\bar{\beta}}^A_0+
{1 \over g_{00}^2}(\kappa - A_0){\bar{\beta}}_{00}^g,	\eqno(3.7)
$$
$$
{\bar{\beta}}^{\tilde{A}}_{i} = {\bar{\beta}}^A_i-
{1 \over g_{00}}\bigr(
(\kappa - A_0)({\bar{\beta}}_{0i}^g+{\bar{\beta}}_{0i}^b)-
(g_{0i}+b_{0i}){\bar{\beta}}^A_0 \bigl)+
{1 \over g_{00}^2}(g_{0i}+b_{0i})(\kappa-A_0)
{\bar{\beta}}_{00}^g,	\eqno(3.8)
$$
where we have used the notation ${\bar{\beta}}^{\tilde{A}}_\mu\equiv
{\bar{\beta}}^A_\mu[\tilde{g},\tilde{b},\tilde{A},\tilde{\phi}]$. 
These are the main equations to be studied in this paper. The task now at 
hand is to see if these two conditions on the gauge field 
$\bar{\beta}$-functions are satisfied by -- and only by -- expressions 
(3.3), (3.6); and if so under what conditions are they satisfied. For that 
we need to perform a standard Kaluza-Klein decomposition of the target 
tensors. This procedure is familiar from previous work \cite{haag,HO,HOS}, 
and in particular we will use the formulas in the Appendix of \cite{HO}, 
supplemented with the ones in the Appendix of this paper.

A final ingredient to such an investigation is the following \cite{freed,sen}. 
At loop order $\ell$, the possible (target) tensor structures 
$T_{\mu\nu...}$ appearing in the sigma model beta functions must scale as 
$T_{\mu\nu...}(\Lambda g^i)=\Lambda^{1-\ell}T_{\mu\nu...}(g^i)$ 
under global scalings of the background fields. In our case at one 
loop, order ${\cal O}(\alpha')$, we have $\ell=1$. These tensor 
structures must obviously also share the tensor properties of the beta 
functions. In our case the gauge beta function is gauge covariant 
Lie algebra valued, with one lower tensor index.

\sect{Duality, the Gauge Beta Function and Heterotic Anomalies}

Let us now start analyzing our main equations, (3.7) and (3.8), for the 
case of the heterotic sigma model, as described in section 2. As 
previously mentioned this model has chiral fermions that, when rotated, 
introduce potential anomalies into the theory. These anomalies need to be 
canceled if the dualization is to be consistent at the quantum level. 
However, our strategy in here is to see if we can get any information on 
this anomaly cancelation from our consistency conditions (3.7-8). So, we will 
set this question aside for a moment and directly ask: are the consistency 
conditions (3.7-8) verified by (3.3), (3.6)? 

We choose to start with torsionless backgrounds. Such choice can be seen 
to extremely simplify equation (3.8), as the metric is parameterized by:
$$
g_{\mu\nu}=
\pmatrix{
a & 0 \cr
0 & \bar{g}_{ij} \cr
}, 	\eqno(4.1)
$$
and we take $b_{\mu\nu}=0$. Therefore, there is also no torsion 
in the dual background \cite{HO,HOS}. In this simpler set up it shall be 
clearer how to deal with anomalies before addressing the case of 
torsionfull backgrounds (see section 5). All this said, equations (3.7-8) 
become,
$$
{\bar{\beta}}^{\tilde{A}}_{0} = {1 \over a}{\bar{\beta}}^A_0+
{1 \over a^2}(\kappa-A_0)\,{\bar{\beta}}_{00}^g,	\eqno(4.2)
$$
$$
{\bar{\beta}}^{\tilde{A}}_{i} = {\bar{\beta}}^A_i. \eqno(4.3)
$$

Now, use the Kaluza-Klein tensor decomposition of (3.3), (3.6), under (4.1), 
and compute ${\bar{\beta}}^{\tilde{A}}_0$ and ${\bar{\beta}}^{\tilde{A}}_i$ 
(see the Appendix). By this we mean the following. One should start with 
(3.3), (3.6), and decompose it according to the parameterization (4.1). 
We will obtain expressions for $\bar{\beta}^A_0$ and $\bar{\beta}^A_i$. Then, 
dualize these two expressions by dualizing the fields according to 
the rules (2.7-9) and (2.11). This yields expressions for 
$\bar{\beta}^{\tilde{A}}_0$ and $\bar{\beta}^{\tilde{A}}_i$. Finally, one 
should manipulate the obtained expressions so that the result looks 
as much as possible as a ``covariant vector transformation'' (1.5). 
Hopefully one would obtain (4.2-3), if the gauge beta functions are 
to satisfy the consistency conditions. However, the result obtained is:
$$
{\bar{\beta}}^{\tilde{A}}_{0}={1 \over a}{\bar{\beta}}^A_0+
{1 \over a^2}(\kappa - A_0)(-{\bar{\beta}}_{00}^g),	\eqno(4.4)
$$
$$
{\bar{\beta}}^{\tilde{A}}_{i}={\bar{\beta}}^A_i.	\eqno(4.5)
$$

The first thing we observe is that even though (4.5) is correct as we 
compare it to (4.3), (4.4) is not as we compare it to (4.2). There is an 
extra minus sign that should not be there. Could anything be wrong? A 
possibility that comes to mind is that nothing is wrong, and indeed (4.4) 
and (4.5) are the correct consistency conditions, implying that the 
duality transformations were incorrect to start with. In that case the duality 
transformation (2.8) would need to be modified in order to yield the correct 
consistency condition upon differentiation. Let us regard this consistency 
condition (4.4) as a differential relation: a one-form 
${\bar{\beta}}^{\tilde{A}}_{0}$ which is expressed in the one-form coordinate 
basis of a ``2-manifold'' with local coordinates $\{A_0,a\}$. But then, as,
$$
{\partial\over\partial a}\,[{1\over a}]=-{1\over a^2} 
\qquad \not\equiv \qquad 
{1\over a^2}={\partial\over\partial A_0}\,[-{1\over a^2}(\kappa-A_0)], 
\eqno(4.6)
$$
we see that the consistency condition (4.4) is {\it not} integrable. 
Therefore we cannot modify the duality transformation rules.

Let us look at this situation from another perspective. We can make 
the consistency conditions (4.4-5) match (4.2-3) if we realize that what 
(4.4) is saying is that, in order for duality to survive as a quantum 
symmetry of the heterotic sigma model, we need to have,
$$
(\kappa - A_0)\,{\bar{\beta}}_{00}^g=0.	\eqno(4.7)
$$
We shall see that this is just the requirement of anomaly cancelation, 
in a somewhat disguised form -- it is the way duality finds to say that 
these anomalies must be canceled, if the dualization is to be consistent 
at the quantum level.

As was mentioned before, equations (2.8-9) were obtained using classical 
manipulations alone. In general, however, there will be anomalies and in 
this case the original theory and its dual will not be equivalent. If we 
want the two theories to be equivalent one must find the required conditions 
on the target fields that make these anomalies cancel. The simplest way to 
do so is to assume that the spin and gauge connections match in the original 
theory, {\it i.e.}, $\omega=A$ \cite{hullwitt,alv1,alv2,gross}. Under this 
assumption, the duality transformation then guarantees that in the dual theory 
spin and gauge connections also match, $\tilde{\omega}=\tilde{A}$. In the 
following we choose to cancel the anomalies according to such prescription.

There are two outcomes of such choice \cite{alv1,alv2}. The first 
one is that if the original theory is conformally invariant to 
${\cal O}(\alpha')$, so is the dual theory. For the sigma model this means 
that flowing to a fixed point will be equivalent to dual flowing to the dual 
fixed point (observe that the duality operation (1.1) does map fixed points 
to fixed points). The second is that we are now required to have 
$\mu=\Omega$, where we define:
$$
\Omega_{\mu\nu}\equiv{1\over2}(\nabla_\mu\xi_\nu-\nabla_\nu\xi_\mu), 
\eqno(4.8)
$$
with $\xi$ the Killing vector generating the Abelian isometry and $\nabla_\mu$ 
the metric covariant derivative. In particular for our adapted coordinates 
$\xi_\mu=g_{\mu0}$, and as the affine connection is metric compatible, 
$\Omega=0$. But then,
$$
\mu_{IJ}=(\kappa-A_0)_{IJ}=0,	\eqno(4.9)
$$
and we are back to (4.7). Then, the consistency conditions are satisfied as 
long as the anomalies are canceled.

Putting together the information in (4.7) and (4.9), let us address a few 
questions. The first thing we notice is that $\bar{\beta}_0^A=0$ as 
$\kappa=A_0$ (recall that in adapted coordinates $\kappa$ satisfies (2.6), 
and so $F_{0i}=0$), which is consistent with the fact that the target gauge 
transformation parameter is not renormalized. Then, the consistency 
conditions become,
$$
{\bar{\beta}}^{\tilde{A}}_0=0 \qquad , \qquad 
{\bar{\beta}}^{\tilde{A}}_i={\bar{\beta}}^A_i,	\eqno(4.10)
$$
stating that the gauge beta function is self-dual under (2.8-9). But so, by 
(4.4-5) with (4.7) satisfied, this proves that (3.6) explicitly satisfies the 
consistency conditions (4.10) -- to the one loop, order ${\cal O}(\alpha')$, 
we are working to.

Given that the gauge field $\bar{\beta}$-function satisfies the consistency 
conditions, the question that follows is whether the scaling arguments 
mentioned in section 3 joined with the consistency conditions (4.10) are 
enough information to uniquely determine (3.3). This would mean that (4.10) 
is verified by, and only by, the correct gauge RG flows of the heterotic 
sigma model. Replacing (3.6) in (4.10) and using the duality transformations, 
we obtain the beta function constraint:
$$
{\beta}^{\tilde{A}}_i={\beta}^A_i+{1\over2}{F_i}^k\partial_k\ln a. 
\eqno(4.11)
$$
On the other hand, according to scaling arguments the possible tensor 
structures appearing in the one loop, order ${\cal O}(\alpha')$, gauge beta 
function are:
$$
\beta^{A}_{\mu} = c_1\,{\bf D}^{\lambda}F_{\lambda\mu}+
c_2\,{H_{\mu}}^{\lambda\rho}F_{\lambda\rho},	\eqno(4.12)
$$
where the notation is as in (3.3). Dealing with torsionless backgrounds (4.1) 
we set $c_2=0$, and are left with $c_1$ alone. Inserting (4.12) in (4.11) 
then yields,
$$
(c_1-{1\over2}){F_i}^k\partial_k\ln a=0,	\eqno(4.13)
$$
and as the background is general (though torsionless), we obtain 
$c_1={1\over2}$ which is the correct result (3.3). Therefore, our consistency 
conditions were able to uniquely determine the one loop gauge field beta 
function, in this particular case of vanishing torsion. We shall later see 
that the same situation happens when one deals with torsionfull backgrounds.

A final point to observe is that the proof of $\mu_{IJ}=0$ through (4.7) 
(and so, also the proof of validity of the consistency conditions) is 
telling us that only if the sigma model is consistent at the quantum level (no 
anomalies) can the duality symmetry be consistent at the quantum level (by 
having the consistency conditions verified). Still, one could argue that 
strictly speaking (4.7) requires either $\mu_{IJ}=0$ or 
$\bar{\beta}^g_{00}=0$. But we also need to cancel all anomalies in order 
to have an RG flow. So, if one wants to flow away from the fixed point along 
all directions in the parameter space, one needs to cancel the anomalies in 
such a way that $\mu_{IJ}=0$ in the adapted coordinates to the Abelian 
isometry. Otherwise, if we were to choose an anomaly cancelation procedure 
yielding non-vanishing $\mu_{IJ}$, it would seem that in order to preserve 
$T$-duality at the quantum level away from criticality, expression 
(4.7) would require that one could only flow away from the fixed point along 
specific regions of the parameter space ({\it i.e.}, regions with 
$\bar{\beta}^g_{00}=0$). As we shall see next when we deal with torsionfull 
backgrounds, this is actually not a good option: the only reasonable choice 
one can make is $\mu_{IJ}=0$.

\sect{Torsionfull Backgrounds}

To complete our analysis, we are left with the inclusion of torsion to the 
previous results. We shall see that even though the calculations are rather 
involved, the results are basically the same. Let us consider the same 
situation as in the last section, with the added flavor of torsion. As in 
\cite{haag,HO}, we decompose the generic metric $g_{\mu\nu}$ as:
$$
g_{\mu\nu}=
\pmatrix{
a & a v_i \cr
a v_i & \bar{g}_{ij}+a v_i v_j \cr
},	\eqno(5.1)
$$
so that $g_{00}=a$, $g_{0i}=a v_i$ and $g_{ij}=\bar{g}_{ij}+av_iv_j$. The 
components of the antisymmetric tensor are written as $b_{0i}\equiv w_i$ 
and $b_{ij}$. We will also find convenient to define the following quantities, 
$a_i\equiv\partial_i\ln a$, $f_{ij}\equiv\partial_iv_j-\partial_jv_i$ and 
$G_{ij}\equiv\partial_iw_j-\partial_jw_i$. From (2.7) one finds that in terms 
of the mentioned decomposition, the dual metric and antisymmetric tensor 
are given by the substitutions $a\rightarrow 1/a$, $v_i \leftrightarrow w_i$, 
and $\tilde{b}_{ij}=b_{ij}+w_iv_j-w_jv_i$.

With all these definitions at hand, we proceed with the Kaluza-Klein 
decomposition of (3.3), (3.6), and compute $\bar{\beta}^{\tilde{A}}_0$ 
(see the Appendix). From the discussion in section 4 it should be clear 
what we mean by this, and which are the several steps required to carry out 
such calculation. Again, one hopes to find (3.7) if the gauge beta function 
is to satisfy the consistency conditions in this torsionfull case. Yet again, 
this does not happen. Instead we obtain,
$$
\bar{\beta}^{\tilde{A}}_0 = {1\over a}\bar{\beta}^A_0+{1\over a^2}
(\kappa-A_0)(-\bar{\beta}^g_{00})+
$$
$$
+{1\over2}(\kappa-A_0)(f^{ij}+
{1\over a}G^{ij})f_{ij}+(f^{ij}+{1\over a}G^{ij})v_iF_{j0}+
{1\over a}v^i[A_0,F_{i0}].	\eqno(5.2)
$$

At first this looks like a complicated result. However, we already have the 
experience from the torsionless case, and that should be enough information 
to guide our way. Indeed, recall the discussion on anomaly cancelation from 
section 4, and proceed to cancel the anomalies according to $\mu_{IJ}=0$. 
Then one has $\kappa=A_0$, and as $\kappa$ satisfies (2.6) in these adapted 
coordinates we are working in, we also have $F_{i0}=0$. Looking again at 
(5.2), one sees that the anomaly cancelation condition -- just like in the 
torsionless case -- makes (5.2) match the consistency condition (3.7). 
Moreover, we also see from (5.2) that, unless we are to severely restrict 
the background fields, the only choice one can make in order to have the 
consistency conditions verified is to cancel the anomalies through 
$\mu_{IJ}=(\kappa-A_0)_{IJ}=0$. Finally, observe that as the target gauge 
transformation parameter does not get renormalized, and $\kappa=A_0$, we will 
have in adapted coordinates $\bar{\beta}^A_0=0$.

One is now left with the analysis of $\bar{\beta}^{\tilde{A}}_i$. Making use 
of all that has been said in the last paragraph this turns out to be a 
reasonable calculation as the consistency conditions (3.7-8) have once 
again become,
$$
{\bar{\beta}}^{\tilde{A}}_0=0 \qquad , \qquad 
{\bar{\beta}}^{\tilde{A}}_i={\bar{\beta}}^A_i,	\eqno(5.3)
$$
the same as (4.10). Computing $\bar{\beta}^{\tilde{A}}_i$ by the usual 
procedure (also see the Appendix), one then finds that it indeed satisfies 
the consistency conditions, modulo gauge transformations. This is reminiscent 
of the fact that the $\beta$-functions only satisfy the consistency conditions 
modulo target reparameterizations, as they are not invariant under such 
transformations. In here, the $\bar{\beta}$-functions themselves are not 
gauge invariant but gauge covariant. In particular, we can choose a gauge 
where the consistency conditions are explicitly verified, the gauge $A_0=0$.

So the gauge field $\bar{\beta}$-function satisfies the consistency 
conditions in the torsionfull case as well as it does in the torsionless 
case. One final question remains: are these consistency conditions enough 
information to compute the coefficients $c_1$ and $c_2$ in (4.12)? The 
constraint these conditions impose on the beta function is obviously the 
same as (4.11). So, when we insert (4.12) in (4.11) we obtain on one hand, 
(4.13). This is to be expected and allows us to determine $c_1={1\over2}$. 
On the other hand we get the new relation,
$$
{1\over2}(c_1-2c_2)(av_i+w_i)f^{jk}F_{jk}=0,	\eqno(5.4)
$$
and as the background is general, we obtain $c_2={1\over4}$ which is the 
correct result (3.3). Therefore, our consistency conditions were able 
to uniquely determine the one loop gauge field beta function. Thus, the 
consistency conditions (5.3) are verified by, and only by, the correct 
RG flows of the heterotic sigma model. In other words, classical target 
space duality symmetry survives as a valid quantum symmetry of the heterotic 
sigma model.

\sect{Conclusions}

We have studied in this paper the consistency between RG flows and 
$T$-duality in the $d=2$ heterotic sigma model. The basic statement 
$[T,R]=0$ that had been previously studied in bosonic sigma models was 
shown to keep its full validity in this new situation, with the added 
bonus of giving us extra information on how one should cancel the anomalies 
(arising from chiral fermion rotations) of the heterotic sigma model. 
Moreover, contrary to previously considered cases \cite{haag,HO,HOS}, 
the requirement $[T,R]=0$ enabled us to uniquely determine the (gauge field) 
beta function at one loop order, without any overall global constant left 
to be determined.

Having considered the cases of closed bosonic, heterotic and (to a certain 
extent) Type II strings/sigma models, a question that comes to mind is the 
following. What happens in the open string case? In the open string case, 
the duality transformations are \cite{dorn1,dorn2},
$$
\tilde{A}_0=0 \qquad , \qquad \tilde{A}_i=A_i.	\eqno(6.1)
$$
The consistency conditions associated to (6.1) are,
$$
\bar{\beta}^{\tilde{A}}_0=0 \qquad , \qquad 
\bar{\beta}^{\tilde{A}}_i=\bar{\beta}^A_i,	\eqno(6.2)
$$
the same as (4.10). Again, using scaling arguments the only possible form 
of the gauge field beta function is (4.12). If actually the Weyl anomaly 
for this situation is the same as in (3.6), we conclude that also in here 
$c_{1}={1\over2}$ and $c_{2}={1\over4}$. Then, by the same line of arguments 
as in section 5, we also conclude that for the open string the statement 
$[T,R]=0$ is true and determines the beta function exactly, ensuring that 
duality is a quantum symmetry of the sigma model. 

One last sigma model to mention is a truncated version of the heterotic sigma 
model \cite{berg1}, where one gets rid of the $\lambda^\mu$ fermions 
in the Lagrangian (2.4). The consequences of such truncation are the loss of 
fermionic partners for the left moving bosons (thus destroying the (1,0) 
supersymmetry), and the fact that one no longer needs to rotate the $\psi^I$ 
fermions in the dualization procedure (thus removing the $\kappa$ parameter 
from expressions (2.8-9) and (3.7-8)). Considering the simpler case of 
torsionless backgrounds, one finds that the known $\bar{\beta}$-functions 
(3.6) satisfy the consistency conditions, modulo gauge transformations. 
Choosing a gauge where these conditions are explicitly verified ($A_0=0$), 
and following the standard dualization procedure \cite{berg1} one obtains 
that the gauge fixed duality transformations are the same as (6.1), and so 
the consistency conditions are the same as (4.10) or (6.2). Then, by the 
familiar line of arguments, $[T,R]=0$ is true and determines the beta function 
exactly ensuring that duality is a quantum symmetry of this sigma model.

Such a basic statement $[T,R]=0$ has now been shown to be alive and well in 
a wide variety of situations, possibly validating the claim in 
\cite{HO,haag2} that it should be a more fundamental feature of the models 
in question than the invariance of the string background effective action.

\vs{5}
\noindent
{\bf Acknowledgments:}
We would like to thank Poul Damgaard, Peter Haagensen and Daniela Zanon for 
comments and reading of the manuscript. R.S. would also like to thank 
S. Bellucci and A.A. Tseytlin for comments/correspondence. K.O. would like to 
thank Harvard University, Department of Physics, for hospitality. R.S. is 
partially supported by the Praxis XXI grant BD-3372/94 (Portugal).

\appendix
\sect{Kaluza-Klein Tensor Decompositions}

We list below all quantities relevant for our computations, as they were 
cited upon during the text. We shall consider in here the general torsionfull 
metric parameterization, as was done in section 5 (see expression (5.1) and 
the definitions that follow it in the text). To use these decompositions in 
section 4, all one needs to do is to set $v_i=w_i=b_{ij}=f_{ij}=G_{ij}=0$ in 
the following. The tensor decompositions are as follows, for both the gauge 
beta function, (3.3), (3.6), and the $(00)$-component of the metric beta 
function, (3.1), (3.4) (where barred quantities will refer to the metric 
${\bar{g}}_{ij}$). Observe that expressions (A.2), (A.4), (A.6) and (A.8) 
below have $(\kappa-A_0)_{IJ}=0$.

\medskip

\noindent 
{\bf 1.} $\nabla^\lambda F_{\lambda\mu}$:
$$
\nabla^\lambda F_{\lambda 0} = {\bar{g}}^{ij}(\partial_iF_{j0}-
{\bar{\Gamma}}^k_{ij}F_{k0})-{1\over2}a_i{F^i}_0-{1\over2}af^{ij}F_{ij}-
av_if^{ij}F_{j0}, 
\eqno({\rm A}.1)
$$
$$
\nabla^\lambda F_{\lambda i} = {\bar{g}}^{jk}(\partial_jF_{ki}-
{\bar{\Gamma}}^\ell_{jk}F_{\ell i}-{\bar{\Gamma}}^\ell_{ji}F_{k\ell})
+{1\over2}a_k{F^k}_i-{1\over2}av_if^{jk}F_{jk}, 
\eqno({\rm A}.2)
$$

\medskip

\noindent
{\bf 2.} $[A^\lambda,F_{\lambda\mu}]$:
$$
[A^\lambda,F_{\lambda 0}] = -v^i[A_0,F_{i0}]+[A^i,F_{i0}], 
\eqno({\rm A}.3)
$$
$$
[A^\lambda,F_{\lambda i}] = -v^j[A_0,F_{ji}]+[A^j,F_{ji}], 
\eqno({\rm A}.4)
$$

\medskip

\noindent
{\bf 3.} ${H_{\mu}}^{\lambda\rho}F_{\lambda\rho}$:
$$
{H_{0}}^{\lambda\rho}F_{\lambda\rho} = 2v_{i}G^{ij}F_{0j}
-G^{ij}F_{ij},	\eqno({\rm A}.5)
$$
$$
{H_{i}}^{\lambda\rho}F_{\lambda\rho}= 2v_{j}G_{ik}F^{kj}
+H_{ijk}F^{jk}, 	\eqno({\rm A}.6)
$$

\medskip

\noindent
{\bf 4.} ${F_\mu}^\lambda\partial_\lambda\phi$:
$$
{F_0}^\lambda\partial_\lambda\phi = {F_0}^i\partial_i\phi, 
\eqno({\rm A}.7)
$$
$$
{F_i}^\lambda\partial_\lambda\phi = {F_i}^j\partial_j\phi, 
\eqno({\rm A}.8)
$$

\medskip

\noindent
{\bf 5.} $\bar{\beta}^{g}_{\mu\nu}$:
$$
\bar{\beta}^g_{00}=-{a\over2}[\bar{\nabla}^i a_i + {1\over2}a_i a^i 
- {a\over2}f_{ij}f^{ij}] - {1\over4} G_{ij}G^{ij} + aa^i\partial_i\phi. 
\eqno({\rm A}.9)
$$

\medskip

Finally one should also include, for the sake of completeness, the 
duality transformations acting on the gauge field strength tensor. These 
are derived directly from expressions (2.8-9), with the following 
results:

\medskip

\noindent
{\bf 6.} $(0i)$-component:
$$
{\tilde{F}}_{0i} = {1\over a} ( F_{0i} - (\kappa-A_0)a_i ), 
\eqno({\rm A}.10)
$$

\medskip

\noindent
{\bf 7.} $(ij)$-component:
$$
{\tilde{F}}_{ij} = F_{ij} - (v_j+{1\over a}w_j) F_{0i} + 
(v_i+{1\over a}w_i) F_{0j} - (\kappa-A_0)(f_{ij} + {1\over a}G_{ij} +
{1\over a}(w_i a_j - w_j a_i)). 
\eqno({\rm A}.11)
$$

\newcommand{\NP}[1]{Nucl.\ Phys.\ {\bf #1}}
\newcommand{\NPPS}[1]{Nucl.\ Phys.\ Proc.\ Suppl.\ {\bf #1}}
\newcommand{\PL}[1]{Phys.\ Lett.\ {\bf #1}}
\newcommand{\CMP}[1]{Comm.\ Math.\ Phys.\ {\bf #1}}
\newcommand{\PR}[1]{Phys.\ Rev.\ {\bf #1}}
\newcommand{\PRL}[1]{Phys.\ Rev.\ Lett.\ {\bf #1}}
\newcommand{\PTP}[1]{Prog.\ Theor.\ Phys.\ {\bf #1}}
\newcommand{\PTPS}[1]{Prog.\ Theor.\ Phys.\ Suppl.\ {\bf #1}}
\newcommand{\MPL}[1]{Mod.\ Phys.\ Lett.\ {\bf #1}}
\newcommand{\IJMP}[1]{Int.\ Jour.\ Mod.\ Phys.\ {\bf #1}}
\newcommand{\JP}[1]{Jour.\ Phys.\ {\bf #1}}
\newcommand{\JMP}[1]{Jour.\ Math.\ Phys.\ {\bf #1}}
\newcommand{\AP}[1]{Ann.\ Phys.\ {\bf #1}}
\newcommand{\CQG}[1]{Class.\ Quant.\ Grav.\ {\bf #1}}
\newcommand{\PREP}[1]{Phys.\ Rept.\ {\bf #1}}


\begin{thebibliography}{99}

\bibitem{porra} A. Giveon, M. Porrati and E. Rabinovici, {\it Target 
Space Duality in String Theory}, \PREP{244} (1994) 77-202, 
{\tt hep-th/9401139}.

\bibitem{lut} C.A. L\"utken, {\it Geometry of Renormalization Group 
Flows Constrained by Discrete Global Symmetries}, \NP{B396} (1993) 
670-692.

\bibitem{bus1} T.H. Buscher, {\it A Symmetry of the String 
Background Field Equations}, \PL{B194} (1987) 59-62.

\bibitem{bus2} T.H. Buscher, {\it Path-Integral Derivation 
of Quantum Duality in Nonlinear Sigma-Models}, 
\PL{B201} (1988) 466-472.

\bibitem{fried} D.H. Friedan, {\it Nonlinear Models in $2+\epsilon$ 
Dimensions}, \AP{163} (1985) 318-419.

\bibitem{freed} L. Alvarez-Gaum\'e, D.Z. Freedman and S. Mukhi, 
{\it The Background Field Method and the Ultraviolet Structure 
of the Supersymmetric Nonlinear $\sigma$-Model}, \AP{134} (1981) 85-109.

\bibitem{haag} P.E. Haagensen, {\it Duality Transformations Away From
Conformal Points}, \PL{B382} (1996) 356-362, {\tt hep-th/9604136}.

\bibitem{HO} P.E. Haagensen and K. Olsen, {\it T-Duality and Two-Loop
Renormalization Flows}, \NP{B504} (1997) 326-342, {\tt hep-th/9704157}.

\bibitem{HOS} P.E. Haagensen, K. Olsen and R. Schiappa, {\it Two-Loop
Beta Functions without Feynman Diagrams}, \PRL{79} (1997) 
3573-3576, {\tt hep-th/9705105}.

\bibitem{haag2} P.E. Haagensen, {\it Duality and the Renormalization 
Group}, in the Proceedings of the NATO Workshop {\sl New Trends 
in Quantum Field Theory}, Zakopane, Poland, 14-21 June 1997, 
ed. P.H. Damgaard and J. Jurkiewicz (Plenum Press, 1997), {\tt hep-th/9708110}.

\bibitem{susy1} B. Zumino, {\it Supersymmetry and K\"ahler Manifolds}, 
\PL{B87} (1979) 203-206.

\bibitem{susy2} L. Alvarez-Gaum\'e and D.Z. Freedman, {\it K\"ahler 
Geometry and the Renormalization of Supersymmetric Sigma Models}, 
\PR{D22} (1980) 846-853.

\bibitem{hullwitt} C.M. Hull and E. Witten, {\it Supersymmetric Sigma Models 
and the Heterotic String}, \PL{B160} (1985) 398-402.

\bibitem{cond1} P.H. Damgaard and P.E. Haagensen, {\it Constraints 
on Beta Functions from Duality}, \JP{A30} (1997) 4681-4686, 
{\tt cond-mat/9609242}.

\bibitem{cond2} C.P. Burgess and C.A. L\"utken, {\it One-Dimensional 
Flows in the Quantum Hall System}, \NP{B500} (1997) 367-378, 
{\tt cond-mat/9611118}.

\bibitem{sw1} A. Ritz, {\it On the Beta-Function in ${\cal N}$=2 
Supersymmetric Yang-Mills Theory}, {\tt hep-th/9710112}.

\bibitem{sw2} J.I. Latorre and C.A. L\"utken, {\it On RG Potential 
in Yang-Mills Theories}, {\tt hep-th/9711150}.

\bibitem{hull} C.M. Hull, {\it Gauged Heterotic Sigma Models}, \MPL{A9} 
(1994) 161-168, {\tt hep-th/9310135}.

\bibitem{alv1} E. Alvarez, L. Alvarez-Gaum\'e and I. Bakas, {\it
T-Duality and Space-Time Supersymmetry}, \NP{B457} (1995) 3-26, 
{\tt hep-th/9507112}.

\bibitem{alv2} E. Alvarez, L. Alvarez-Gaum\'e and I. Bakas, {\it
Supersymmetry and Dualities}, \NPPS{46} (1996) 16-29, 
{\tt hep-th/9510028}.

\bibitem{gross} D.J. Gross, J.A. Harvey, E. Martinec, and R. Rohm, 
{\it The Heterotic String}, \PRL{54} (1985) 502-505.

\bibitem{bell} S. Bellucci and R.N. Oerter, {\it Weyl Invariance of the 
Green-Schwarz Heterotic Sigma Model}, \NP{B363} (1991) 573-592.

\bibitem{bell2} S. Bellucci, {\it Ultraviolet Finiteness versus 
Conformal Invariance in the Green-Schwarz $\sigma$-Model}, 
\PL{B227} (1989) 61-67.

\bibitem{berg1} E. Bergshoeff, I. Entrop and R. Kallosh, {\it
Exact Duality in String Effective Action}, \PR{D49} (1994) 6663-6673, 
{\tt hep-th/9401025}.

\bibitem{zano} M.T. Grisaru, H. Nishino and D. Zanon, {\it 
$\beta$-Functions for the Green-Schwarz Superstring}, \NP{B314} 
(1989) 363-389.

\bibitem{sen} A. Sen, {\it Equations of Motion for the Heterotic 
String Theory from the Conformal Invariance of the Sigma Model}, 
\PRL{55} (1985) 1846-1849.

\bibitem{grig} G. Grignani and M. Mintchev, {\it The Effect of Gauge
and Lorentz Anomalies on the Beta Functions of the Heterotic Sigma
Model}, \NP{B302} (1988) 330-348.

\bibitem{calla} C.G. Callan, D. Friedan, E.J. Martinec and 
M.J. Perry, {\it Strings in Background Fields}, \NP{B262} (1985) 593-609.

\bibitem{tsey1} A.A. Tseytlin, {\it Conformal Anomaly in a 
Two-Dimensional Sigma Model on a Curved Background and Strings}, 
\PL{B178} (1986) 34-40.

\bibitem{tsey2} A.A. Tseytlin, {\it Sigma Model Weyl Invariance 
Conditions and String Equations of Motion}, \NP{B294} (1987) 383-411.

\bibitem{tsey3} A. Giveon, E. Rabinovici and A.A. Tseytlin, 
{\it Heterotic String Solutions and Coset Conformal Field Theories}, 
\NP{B409} (1993) 339-362, {\tt hep-th/9304155}.

\bibitem{dorn1} H. Dorn and H.J. Otto, {\it Remarks on T-Duality for 
Open Strings}, \NPPS{56B} (1997) 30-35, {\tt hep-th/9702018}.

\bibitem{dorn2} H. Dorn, {\it Nonabelian Gauge Field Dynamics on 
Matrix D-Branes in Curved Space and Two-Dimensional $\sigma$-Models}, 
in the Proceedings of the Workshop {\sl Quantum Aspects of Gauge Theories, 
Supersymmetry and Unification}, Neuchatel, Switzerland, 18-23 Sep 1997, 
{\tt hep-th/9712057}.
 
\end{thebibliography}
\end{document}